\documentclass[twocolumn,showpacs,prb,preprintnumbers,amsmath,amssymb]{revtex4}
\makeatletter
\def\@dotsep{4.5}
\makeatother

\usepackage{epsfig,graphicx}
\usepackage{dcolumn}
\usepackage{bm}
\usepackage{subfigure}

\begin{document}

\preprint{}

\title{A diagrammatic formulation of the kinetic theory of fluctuations in equilibrium classical fluids. VI. Binary collision approximations for the memory function for self correlation functions}

\author{Joyce E. Noah-Vanhoucke}
\author{Hans C. Andersen}%
 \email{hca@stanford.edu}
\affiliation{Department of Chemistry, Stanford University, Stanford, California 94305}

\date{\today}

\begin{abstract}
We use computer simulation results for a dense Lennard-Jones fluid for a range of temperatures to test the accuracy of various 
binary collision approximations for the memory function for density fluctuations in liquids.  The approximations 
tested include the moderate density approximation of the generalized Boltzmann-Enskog memory function (MGBE) of 
Mazenko and Yip, the binary collision approximation (BCA) and the short time approximation (STA) of Ranganathan 
and Andersen, and various other approximations derived by us using diagrammatic methods.  The tests  are of two 
types.  The first is a comparison of the correlation functions predicted by each approximate memory function 
with the simulation results, especially for the self longitudinal current correlation function (SLCC).  The 
second is a direct comparison of each approximate memory function with a memory function numerically extracted 
from the correlation function data.  The MGBE memory function is accurate at short times but decays to zero too 
slowly and gives a poor description of the correlation function at intermediate times.  The BCA is exact at zero 
time, but it predicts a correlation function that diverges at long times.
The STA gives a reasonable description of the SLCC but does not predict the correct temperature dependence of 
the negative dip in the function that is associated with caging at low temperatures.  None of the other binary 
collision approximations is a systematic improvement upon the STA.  The extracted memory functions have a 
rapidly decaying short time part, much like the STA, and a much smaller, more slowly decaying part of the type 
predicted by mode coupling theory. Theories that  use mode coupling commonly include a binary collision term in 
the memory function but do not discuss in detail the nature of that term.  It is clear from the present work that 
the short time part of the memory function has behavior associated with brief binary {\em repulsive} collisions, 
such as those described by the STA.  Collisions that include attractive as well as repulsive interactions, such 
as those of the MGBE, have a much longer duration, and theories that include them have memory functions that 
decay to zero much too slowly to provide a good first approximation for the correlation function.  This 
leads us to speculate that the memory function for density fluctuations can be  usefully regarded as a sum of 
at least three parts: a contribution from repulsive binary collisions (the STA or something similar to it), 
another short time part that is related to all the other interactions (but whose nature is not understood), and 
a longer time slowly decaying part that describes caging (of the type predicted by mode coupling theory).  
\end{abstract}

\maketitle

\section{\label{sct:intro}Introduction}
The concept of uncorrelated binary collisions\cite{note1} has played an important role in the development 
of the kinetic theory of fluids.\cite{note2} It is a central idea in the Boltzmann
kinetic equation for dilute gases, the  Enskog equation for high density liquids, and generalizations of
Enskog theory that take into account the softness of the repulsive
potential.\cite{curtiss64-III,curtiss65-IV,curtiss65-V,Chapman,Hirschfelder} 
The nature of the binary
collisions in the Boltzmann and Enskog theories are very different, involving in one case the entire
interparticle potential, and in the other case only the repulsive part of the potential, which is  idealized
as a hard sphere potential.

For kinetic theories of dense liquids that do take into account the effect of both the attractive and 
repulsive parts of the interatomic potential on the dynamics, the different range, strength, and effects 
of the two parts of the potential has motivated the construction of theories that treat these two parts 
very differently.  Examples include the Rice-Allnatt theory\cite{allnatt61,rice61,sengers61} and the Karkheck-Stell 
theory.\cite{karkheck80,karkheck81,karkheck82,stell83,karkheck85,karkheck88,stell89}

The equilibrium theory of the structure and thermodynamics of liquids shows similar developments. At 
low density, both the radial distribution function and the thermodynamic properties are represented by 
relatively simple expressions that contain the entire potential.  At higher density, various theories 
have invoked the primacy of the repulsive forces for determining the 
structure.\cite{widom64,reiss65,widom67,henderson67,barker67,andersen71,andersen83}
Theories that take into account both types of forces at high density often calculate the effect of 
attractions using very different ideas from those used for repulsive forces.\cite{widom64,lebowitz65,henderson67,barker67,andersen72}

Theories of equilibrium structure and thermodynamics have been derived and formulated in a variety of ways,
including perturbation theory, integral equations, and cluster expansions.\cite{hansen-mcdonald,mayer76}  
The Mayer cluster theory and its extensions\cite{mayer76,morita61,stell64,andersen77} provide a
unifying theoretical framework for deriving and analyzing many of these theories (and many others) and 
relating them to one another. The Mayer theory represents the static correlation functions and free energy 
of a fluid in terms of an infinite set of diagrams.  Approximations are typically developed by identifying 
certain subsets of diagrams, summing them, and discarding the others. A standard tactic in  
cluster theory is to decompose the objects in the theory that represent interactions into various pieces 
and include the various pieces in different ways in developing 
approximations.\cite{andersen71,andersen77,andersen83,stell64}

A focus on interparticle forces and their effects on dynamics is quite different from the perspective
provided by the fully renormalized kinetic theory of Mazenko and 
co-workers.\cite{mazenko:72,mazenko:73I,mazenko:73II,mazenko:74III,mazenko77}
In the latter approach, the theory of the correlation functions for density fluctuations
in equilibrium liquids is formulated in such a way that the interparticle potential does not appear;
instead, only the equilibrium static correlation functions, such as the pair correlation function and its
generalization to more than two particles, appear in the final results.  In effect, the dynamics is
expressed in terms of the potentials of mean force rather than the `bare' potential.

One of the results of this theory was a generalized Boltzmann-Enskog memory
function (denoted GBE) for fluids and an approximation to this function that they called the 
`moderate density approximation' to the generalized Boltzmann-Enskog memory function (denoted MGBE).
The latter can be interpreted physically as taking into account 
uncorrelated binary collisions among the particles, with the potential that determines the dynamics being 
the potential of mean force rather than the bare potential.

In recent work, Young and Andersen\cite{tomMDI,tomMDII} used molecular dynamics simulations to study
the behavior of several pairs of atomic liquids.  Each pair consisted of two systems at the same
density but different temperatures, with very similar pair correlation functions, and hence very similar
potentials of mean force, despite their very different interparticle potentials.  It was found that
some (but not all) of the features of the dynamics of such a pair of liquids were very similar,
suggesting that the potential of mean force is in some sense a more important determinant of
dynamics than is the bare potential itself.  This is consistent with the basic ideas of renormalized
kinetic theory.

A formally exact diagrammatic formulation of the kinetic theory of density correlation functions was 
developed by one of us.\cite{andersen:dkt1,andersen:dkt2,andersen:dkt3}
It expresses the dynamic correlation functions of the
particle density in terms of an infinite series of diagrams in a way analogous to the infinite series in 
the Mayer cluster theory for equilibrium static correlation functions. The vertices of the
graphs, which represent interactions among particles, are expressed in terms of static correlation functions, 
and the theory is fully renormalized in the sense of Mazenko. This theory can provide a unifying theoretical 
framework for deriving and analyzing kinetic theories of density fluctuations in liquids.

Using this graphical theory, Ranganathan and Andersen\cite{andersen:dkt4,andersen:dkt5} 
considered the case of a fluid whose interatomic potential contained a continuous but very repulsive short 
ranged part, as well as longer ranged attractions. Using the inverse of the strength of the short ranged 
force as a small parameter, they determined the diagrams that are most important for the short time behavior 
of the memory functions for the dynamic correlation function for density fluctuations. This analysis involved 
a procedure much like the one discussed above for the Mayer theory, namely a decomposition of the fully 
renormalized quantities that appear in the diagrams of the theory into various contributions with different 
characteristics in the limit as the small parameter goes to zero. The most important diagrams 
({\it i.e.} the most divergent diagrams) were identified. Summing this infinite series of diagrams exactly led 
to an approximation for the memory function that is called the short time approximation (STA).  
The structure of this approximation indicates that it takes into account uncorrelated binary 
collisions involving the repulsive part of the potential, with a collision frequency that is determined 
by the structure of the fluid (and hence is influenced by the attractive forces).  

Using this short time approximation and neglecting the longer time contributions to the memory function gives
a kinetic theory that can be regarded as a generalization of the Enskog theory to dense fluids with continuous
repulsive forces as well as attractive forces.  The Enskog kinetic equation is local in time, with no memory 
of the past history. As a result, the memory function in the Enskog theory is proportional to a Dirac 
delta function in time. For fluids with continuous repulsive potentials, there is a part of the memory 
function that is very large at very short times\cite{verlet} and that approaches a delta function in 
time as the repulsive part of the potential approaches a hard sphere potential. The STA includes all 
diagrammatic contributions to that large, short-lived part of the memory function. This is the sense in 
which the STA is a generalization of the Enskog theory.         

The predictions of this theory have been tested by comparison with computer simulation studies of the 
Lennard-Jones fluid.\cite{andersen:dkt5} The accuracy of the theory is different for different correlation 
functions but is surprisingly good at high temperatures for the viscosity and diffusion constant.  
The theory fails to describe the changes in correlation functions that take place as a dense high 
temperature fluid is cooled to the triple point temperature.  In particular, the increasingly negative dip 
in the velocity autocorrelation function and the self-longitudinal current are not well described, 
suggesting that the dynamics in the vicinity of the triple point requires much more than uncorrelated 
repulsive binary collisions for its description.

Ranganathan and Andersen also identified a more general set of diagrams whose sum leads to an approximation, 
called the binary collision approximation (BCA), that describes the effect of uncorrelated binary collisions 
that are determined by the potential of mean force. (The relationship of this to the moderate density 
approximation of the generalized Boltzmann-Enskog memory function will be discussed in more detail below.)

In this paper, we discuss a variety of ways in which binary collisions might be defined for use in
approximations based on uncorrelated binary collisions.  We test several of them by comparing their 
predictions with the results of molecular dynamics simulations of a dense Lennard-Jones liquid at various 
temperatures, from the triple point temperature to four times that value. There were several motivations 
for this work.

\noindent 
1.\ \, There are many different possibilities for specifying the nature of the binary collisions in dense
fluids. Some of them are mentioned above. Others are suggested by the possibility of breaking the 
potential of mean force, which appears in fully renormalized kinetic theory, into various contributions 
and describing the effect of the various contributions in different ways, in analogy to the way in which 
the bare potential is decomposed in both kinetic theory and the equilibrium theory of fluids.
Moreover, the diagrammatic kinetic theory suggests some straightforward extensions of binary collision 
approximations (as is discussed below).

\noindent 
2.\quad It would be worthwhile to test the various approximations by direct comparison of their predictions
with computer simulation studies in order to determine which of them provides, in some sense, 
the best starting point for further development of kinetic theory.

\noindent 
3.\quad The mode coupling theory of G\"otze and coworkers, and the kinetic theory that provides the basis
for it,\cite{sjogren:80I,sjogren:80II,sjogren:80III}
focusses on the long time behavior of the memory function for density 
correlations in a fluid, but it assumes that there is a short time part that represents brief and 
presumably binary collisions between the atoms.  However, the theory does not provide a microscopic 
expression for this short time part, and in practice its contribution is described by a simple empirical 
function of time obtained by fitting to simulation or experimental data. It would be worthwhile to 
understand the nature of the true short time behavior of the memory function as a starting point for 
understanding the approximations made in mode coupling theory and attempting to go beyond those
approximations.

Our discussion will rely heavily on the diagrammatic kinetic 
theory\cite{andersen:dkt1,andersen:dkt2,andersen:dkt3,andersen:dkt4} as a
unifying framework for stating and comparing the various approximations as well as for deriving some
of them.  We restrict attention to dense atomic liquids (rather than gasses) with short range forces
(rather than Coulombic forces).  We also restrict attention to theories that calculate and use a 
memory function to calculate the
equilibrium time correlation functions of the density in phase space.  Finally,
the correlation functions we focus on in this paper are self (or incoherent) correlation functions.

\section{\label{sct:bca}Binary collision approximations and the diagrammatic theory}

\subsection{Self correlation function}
We are concerned with the kinetic theory of the self correlation function of density fluctuations in
a dense atomic fluid.  The incoherent scattering function and self-diffusion are experimental 
observables associated with this function. (For additional studies that deal with other correlation 
functions, see Ref.~\onlinecite{jen:thesis}.)

The system of interest is a single component classical atomic fluid at equilibrium. Its Hamiltonian is
$$H({\bf r}^N,{\bf p}^N)=\sum_{i=1}^N|{\bf p}_i|^2/2m+\sum_{i<j=1}^Nu(|{\bf r}_i-{\bf r}_j|)$$
The self part of the two-point time correlation function for density fluctuations in single particle phase 
space is
$$  C^{(s)}(11',t)\equiv 
\frac{1}{\rho}\left\langle\sum_{i=1}^N \delta f_i(1,t)\delta f_i(1',0)\right\rangle $$
where $\rho=N/V$ is the number density, 
$$ f_i(1,t) = \delta(\mathbf R_1-\mathbf r_i(t))\delta(\mathbf P_1-\mathbf P_i(t)) $$
is the density in single particle phase space for particle $i$, and
$$ \delta f_i(1,t) \equiv f_i(1;t)-\langle f_i(1;t) \rangle$$
is a density fluctuation.
We use numbers like 1, 2, {\it etc.} to represent points in single particle phase space,
{\it e.g.,} $1=\mathbf R_1\mathbf P_1,\ 2' = \mathbf R_{2'}\mathbf P_{2'}$. Angular brackets denote 
equilibrium ensemble averages.

The kinetic equation for this correlation function is
\begin{eqnarray}
  \lefteqn{\frac{\partial}{\partial t}C^{(s)}(11',t)} \nonumber\\
\label{eq:kineticEquation} 
      &=& \int_0^t dt'\,\int d3\, M^{(s)}(13;t-t')C^{(s)}(31',t')
\end{eqnarray}
where the self part of the memory function $M^{(s)}$ is the sum of two terms
$$  M^{(s)}(11',t) = M_f(11',t) + M_c^{(s)}(11',t) $$
The flow term is
$$ M_f(11',t) = -\frac{\mathbf P_1}{m}\cdot\nabla_{\mathbf R}\delta(11')\delta(t) $$
We use the notation $\delta(11') = \delta(\mathbf R_1-\mathbf R_{1'})\delta(\mathbf P_1-\mathbf P_{1'})$ 
and $\nabla_{\mathbf R}\delta(11')$ is the gradient of $\delta (11')$ with respect to the first position 
argument. We make approximations for $M^{(s)}_c$, the collisional part of the self memory function,
which we shall simply refer to as the memory function.

The initial condition for eq.~(\ref{eq:kineticEquation}) is
$$  C^{(s)}(11',0) = \rho M_m(1)\delta(11') $$
where  $M_m$ is the Maxwell-Boltzmann distribution of momenta
$$ M_m(1)\equiv (2\pi mk_BT)^{-3/2}\exp\left(-|\mathbf P_1|^2/2mk_BT\right)$$
and $g(11')\equiv g(|\mathbf R_1~-~\mathbf R_{1'}|)$
is the pair correlation function of the fluid.

\subsection{Diagrammatic series}
The diagrammatic kinetic theory provides several closely related expressions for the correlation 
function $C^{(s)}$ and the memory function $M^{(s)}$ as the sum of the values of an infinite series of 
diagrams of a certain topological structure.\cite{andersen:dkt1,andersen:dkt2,andersen:dkt4}
The diagrams contain propagators that 
describe the time evolution of density fluctuations and vertices that represent renormalized interactions 
among those fluctuations. We shall focus on the two formulations of these series that are most useful for 
the present discussion.

The first form of diagrammatic series\cite{andersen:dkt2} contains $\chi^{(0)}$ propagators, 
which describe density fluctuations at sets of one or more points in the fluid, and $Q$ vertices.

The second form of the diagrammatic series\cite{note5}
contains $\chi^{(fp)}$ 
propagators, each of which describes the propagation of a single density fluctuation associated with a 
single particle that moves according to free particle motion, and $Q^{(c,p)}$ vertices, which describe 
the localized interactions among such density fluctuations. 

The two series are each formally exact. The second is derived from first. The first is useful for proving 
some formal properties of the series, such as symmetry properties. In practice, the second is more useful 
for deriving practical approximations. In the second series it is possible to identify density fluctuations 
in different parts of a diagram and interpret them as being caused by the presence (or absence) of the 
same particle. The vertices in the theory are calculated from static correlation functions of the fluid, 
and so it is possible to identify how many distinct correlated particles are associated with a vertex.
The identification of the number of particles associated with a vertex is in some cases ambiguous, 
for reasons that we will discuss, but it is still possible to identify terms that correspond to, 
for example, the interaction of two particles, and hence it is possible to discuss whether diagrams 
describe binary collisions.

Additional diagrammatic series can be derived from the second series by expressing the vertices as sums 
of various contributions.  For example, in vertices that contain the potential of mean force, the latter 
can be decomposed into a short ranged repulsive part and a longer ranged oscillatory part, leading to 
the original vertex being expressed as a sum of two new vertices.  The resulting diagrammatic series 
can then be used to develop approximations in which the effects of the two parts of the potential are 
taken into account in different ways. (When an approximation is made to a diagrammatic series for a 
correlation function, by neglecting some diagrams and/or approximating certain vertices, the resulting
approximation may or may not be consistent with known symmetry properties of the correlation function,
such as stationarity and time reversal symmetry. See Appendix~\ref{appendix:sym} for a discussion.)
 
\subsection{Diagrammatic characterization of various approximations}
\paragraph{Generalized Boltzmann-Enskog memory function.} 
This memory function, denoted GBE, is generated in Mazenko's theory 
by neglecting a function $\Gamma$, which represents renormalized dynamical interactions, in the equation of 
motion for a function in terms of which the memory function of interest is 
expressed.\cite{note6}
In terms of the first diagrammatic series, it represents the sum of all diagrams in the series for the memory 
function in which there are at most two density fluctuations propagating at any time.

This approximation satisfies important symmetry properties exactly (see Appendix~\ref{appendix:sym}).  
Moreover, it gives exactly the correct value of the 
memory function at $t=0$.  However, it is not clear how it might be calculated in any practical fashion, 
and to our knowledge numerical calculations of the GBE memory function have never been performed. 
The physical meaning of the approximation is hard to specify but the mathematical meaning in the context 
of the diagrammatic theory is clear.  It contains contributions to the memory function that describe 
two density fluctuations that interact with each other and propagate. The identities of the two particles
associated with the two fluctuations can be different at different times.
All interactions and propagations of the two fluctuations are 
treated exactly. The approximation that is made  is to neglect the possibility of three or more simultaneous 
fluctuations.  

\paragraph{Binary collision approximation.}  
This approximation, denoted BCA, was defined 
in Ref.~\onlinecite{andersen:dkt4} in terms of the second diagrammatic series.  
It is the sum of all diagrams in the series for
the GBE memory function in which the same pair of particles is associated with the fluctuations at all times.

This approximation does not satisfy important symmetry properties exactly (see Appendix~\ref{appendix:sym}).  
However, it does give exactly the correct value for $M^{(s)}$ at $t=0$.  A practical algorithm 
exists to evaluate it, as is discussed below.

The physical meaning of the approximation is that it contains contributions to the memory function that 
describe two density fluctuations, associated with two specific particles, that interact with each other 
and propagate, without having three fluctuations present at any time. All interactions and propagations 
of the two fluctuations are treated exactly. The approximation here is not only to neglect the possibility 
of three or more simultaneous fluctuations (as in the GBE) but also to neglect diagrams in which the two 
fluctuations at one time represent a different pair of particles than the two fluctuations at other times. 
This characterization is the reason why it can be regarded as a binary collision approximation, 
{\it i.e.} one that involves only two particles.

\paragraph{The moderate density approximation to the generalized Boltzmann-Enskog memory function.}  
This memory function, denoted MGBE, was defined by Mazenko as an 
approximation to the GBE.  Various quantities in the GBE are defined in terms of static correlation functions 
of the fluid. To construct the MGBE, the contributions to these quantities that contain three and four 
particle correlation functions are deleted.\cite{note7}

The deletion of the three and four particle correlation functions to define the MGBE is closely related to, 
but not equivalent to, what is done in defining the BCA to include diagrams in which fluctuations 
representing at most two distinct particles appear.  (See Appendix~\ref{appendix:leftVertex}.) 
The MGBE is almost 
equivalent to the BCA, except for one factor in every diagram.  From a diagrammatic point of view, 
the MGBE can be regarded as an approximation to the BCA in which the vertex $Q_{12}$ that appears 
in each diagram is replaced by an approximation that contains the potential of mean force. 
(A detailed expression for the MGBE is given below, together with a comparison to the BCA.) 

This approximation satisfies important symmetry properties exactly (see Appendix~\ref{appendix:sym}).
It does not give exactly the correct value of the memory 
function at $t=0$. A practical algorithm exists to evaluate it, as is discussed below.

The physical meaning of the MGBE is equivalent to that of the BCA.  The only difference is that the 
additional approximation made to the $Q_{12}$ leads to a binary collision approximation that satisfies 
the symmetry condition.

\paragraph{Short time approximation.} 
This will be referred to as the STA. This was derived\cite{andersen:dkt4} from the second diagrammatic series 
using the procedure discussed above, namely expansion of the diagrams in powers of  a small parameter 
representing the inverse of the strength of the repulsive part of the potential and retaining those terms 
that at short times are most divergent. The diagrams that appear are those of the BCA, as defined above, 
but with vertices that contain the repulsive part of the interatomic potential, rather than the potential 
or the potential of mean force. (See below for a more precise statement of the value of the STA.)

The STA satisfies the symmetry property approximately, but not exactly. (In the limit of infinitely 
repulsive forces, it satisfies the symmetry property exactly.  See Appendix \ref{appendix:sym}.) 
It does not give 
exactly the correct value of the memory function at $t=0$. A practical algorithm exists to evaluate it.

\paragraph{Other binary collision approximations.}  Below we shall discuss several approximations whose 
diagrammatic meaning is straightforward. Like the STA, they can be defined graphically by starting with 
the BCA, separating each of the vertices into two terms, and retaining only the first term.  
The physical meaning of such an approximation depends on the nature of the separation. The reason for 
investigating such approximations is the desire to develop better binary collision approximations as a 
starting point for a more general kinetic theory.

\paragraph{Beyond the binary collision approximation.}
A major restrictive assumption of all the approximations mentioned above is that the  possibility of 
having three or more simultaneous density fluctuations is ignored. The pair of particles that are associated with 
the two density fluctuations interact with each other, but the presence of other particles is taken into 
account only in some average sense, if at all, due to the appearance of static pair correlation functions 
of the fluid in the various vertices.

\begin{figure}
\epsfig{figure=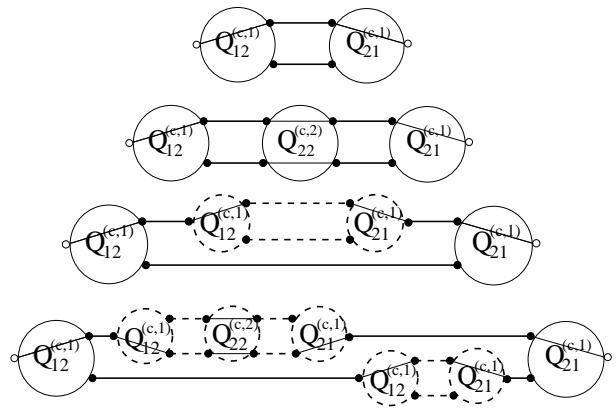,width=8cm}
\caption{The first two diagrams are members of the diagrammatic series for the binary collision
memory functions. (The remaining diagrams in the binary collision memory functions have
additional $Q_{22}^{(c,2)}$ vertices on the two horizontal particle paths.) The third and 
fourth diagrams are examples of diagrams that are not in a binary collision memory function.
One or both of the two particles is interacting separately with a third particle in an STA
approximation. The parts of the diagrams that have these STA interactions are indicated with
dashed lines and dashed circles. (See the text and Fig.~\ref{fig:diagrams-sbca2} for additional
discussion.)}
\label{fig:diagrams-sbca1}
\end{figure}

\begin{figure}
\epsfig{figure=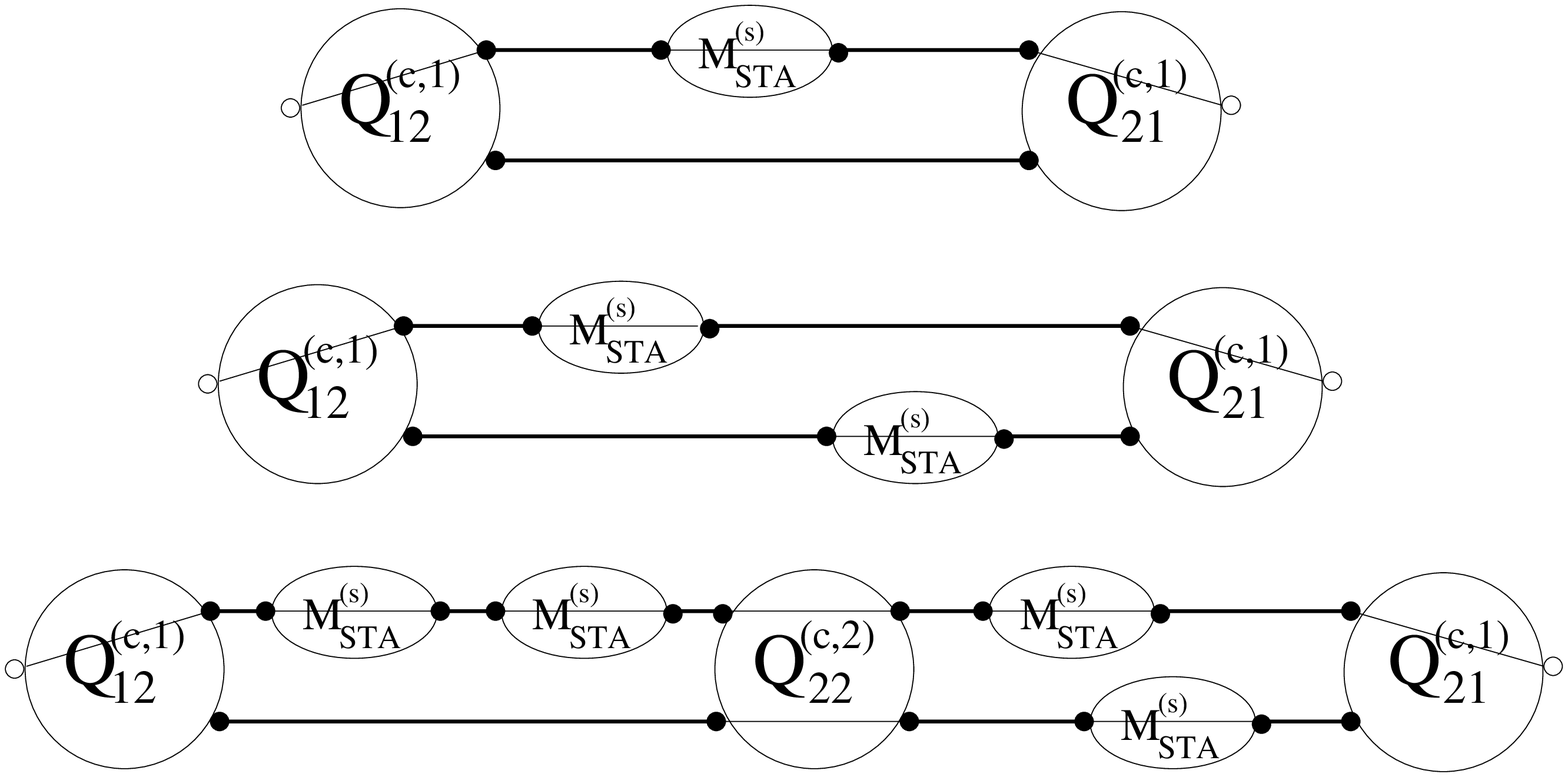,width=8cm}
\caption{Examples of diagrams in a binary collision approximation augmented with $M_{STA}^{(s)}$ memory
functions on the particle paths to represent interactions with third particles of the type shown in 
Fig.~\ref{fig:diagrams-sbca1}. The first diagram here includes the third diagram of 
Fig.~\ref{fig:diagrams-sbca1}. The second diagram here includes the fourth diagram of
Fig.~\ref{fig:diagrams-sbca1}. The generic member of this series is obtained by taking the generic
member of the series for a binary collision approximation (see the first two diagrams in
Fig.~\ref{fig:diagrams-sbca1} for examples) and inserting an arbitrary number of $M_{STA}^{(s)}$ 
memory functions on each of the lines. (See the text and Fig.~\ref{fig:diagrams-sbca1} for 
additional discussion.)}
\label{fig:diagrams-sbca2}
\end{figure}

It is relatively simple, in the context of a graphical theory, to construct approximations that relax these 
assumptions in a physically meaningful way. The first two diagrams in a binary collision approximation are shown
as the first two diagrams in Fig.~\ref{fig:diagrams-sbca1}. The generic BCA diagram consists of two particle
paths\cite{andersen:dkt4} that propagate from the 
$Q_{21}^{(c,1)}$ vertex on the right to the $Q_{12}^{(c,1)}$ vertex on the left. In between these vertices 
are zero or more $Q_{22}^{(c,2)}$ vertices. These diagrams describe the propagation of a pair of particles 
that interact with each other using the potential that is contained within the $Q^{(c,2)}_{22}$ function. 
To develop the $K$ approximations, consider including diagrams in the exact graphical series that are similar 
to these and that are constructed by inserting STA memory functions into the propagators
on the particle paths of the original diagrams of the BCA. The diagrams that result are those 
of the original BCA diagrams with any number of STA memory functions inserted.  
See Fig.~\ref{fig:diagrams-sbca2}.

When the diagrams with STA memory functions are evaluated, the two points have different times associated with 
them and the STA memory function is  a function of the time difference. Since the collisions are brief, we 
approximate the effect of these vertices using a BGK model.\cite{krook} The BGK model memory function is
$$M_K(11^\prime;t) =\nu m_K(11^\prime)\delta(t)$$
where
$m_K(11^\prime) =-\delta({\bf P}_1-{\bf P}_1^\prime)+M_m({\bf P}_1)$
 and $\nu$ is a collision frequency.  The BGK model is an approximate memory function that is instantaneous 
in time. It approximates a collision as having the 
effect of replacing the momentum of the particle by a new random momentum drawn from a Maxwell-Boltzmann 
distribution.  Because of its simplicity, its effect can easily be incorporated into the algorithm for 
evaluating binary collision approximations, as is discussed below.

This procedure allows any binary collision approximation to be converted into one that also
includes brief uncorrelated binary collisions of each of the two particles with third particles or
fluctuations in the surrounding fluid.  Graphically this is done by inserting $M_K$ vertices, as
approximations for STA memory functions, into the $\chi^{(fp)}$ propagators that by themselves
describe free particle propagation.  When this procedure is applied to any approximation, the
resulting approximation will be given a name with ``K'' as a prefix.  Thus, for example, the MGBE
approximation, when modified in this way, will be called the K-MGBE approximation.  In constructing 
such an approximation, the value of the parameter $\nu$ must be chosen.  This will  be discussed below.

We shall refer to this class of approximations as $K$ approximations. $K$ approximations are mathematically
and physically similar to a type of approximation suggested by Mazenko and Yip.\cite{note8}

\subsection{Analytic formula for the BCA and related approximations}
The various approximations mentioned above, with the exception of the GBE approximation, can be expressed 
in terms of very similar analytic formulas for $M^{(s)}$.  As a result, all of them can be evaluated with 
a similar algorithm. Here we will state the formulas. The algorithm will be discussed later.

For the BCA, the memory function is 
\begin{eqnarray}
\label{eq:MBCA-generalForm}
  \lefteqn{M^{(s)}_{BCA}(1t,1't') =\int d2\,d3\,d4\,d5\,}\\
&\times& Q_{12}^{(c,1)}(1;23)\chi_{BCA}(23t;45t') Q_{21}^{(c,1)}(45;1')\nonumber
\end{eqnarray}
where the $Q$ vertices represent interaction terms and the $\chi_{BCA}$ term is a response function.
The detailed equations for these functions are
\begin{eqnarray}
\label{eq:orig_Q12}
 \lefteqn{ Q_{12}^{(c,1)}(1;1'2') = \nabla_{\mathbf R}V_{12}(12')\cdot\nabla_{\mathbf P}\delta(11') }\\
\label{eq:orig_Q21}
  \lefteqn{Q_{21}^{(c,1)}(12;1')}\\ &=& \frac{M_m(1)}{M_m(1')} \rho(2)M_m(2)g(12)
         \nabla_{\mathbf R} V_{21}(12)\cdot \nabla_{\mathbf P}\delta(11')\nonumber\\
\label{eq:orig_chiBCA}
\lefteqn{\chi_{BCA}(12t;1'2't')}\\
 &=& e^{-i\mathcal L(12)(t-t')} \delta(11')\delta(22')\Theta(t-t')\nonumber
\end{eqnarray}
The Liouville operator $\mathcal L$ is
\begin{eqnarray}
\label{eq:fullL}
  \mathcal L(12) &=& L_0(1) + L_0(2) + \tilde L_1(12)\\
\label{eq:L0}
  L_0(1) &=& -i\frac{\mathbf P_1}{m(1)}\cdot\nabla_{\mathbf R_1}\\
\label{eq:L1}
  \tilde L_1(12) &=& i\nabla_{\mathbf R}V_{22}(12)\cdot(\nabla_{\mathbf P_1} -\nabla_{\mathbf P_2})
\end{eqnarray}
These formulas contain three functions $V_{12}$, $V_{21}$ and $V_{22}$ that are two particle potentials.  
For the BCA, they are 
\begin{equation}
\label{eq:allV}
 V_{12}(12)=u(12);\quad V_{21}(12)=V_{22}(12)=v_{MF}(12)
\end{equation}
where $u(12)$ is the interparticle potential and $v_{MF}(12)=-k_BT\ln\,g(12)$ is the potential of mean force, 
and $g(12)$ is the pair correlation function.

The BCA is exact in the low density limit and corresponds to the linearized Boltzmann equation in that 
limit on appropriate length and time scales. The $\chi_{BCA}$ describes two particle scattering with
an effective interparticle potential that is the potential of mean force, which also appears in $Q_{21}$.
At low density, the potential of mean force approaches the bare potential. The BCA describes binary 
collisions (on microscopic length and time scales) in much the same way as the Boltzmann equation, with
some many body effects accounted for in the effective potential. 

The MGBE is of the same form as eqs.~(\ref{eq:MBCA-generalForm})--(\ref{eq:L1}), with (\ref{eq:allV}) 
replaced by 
\begin{equation}
\label{eq:MGBE-allV}
     V_{12}(12)= V_{21}(12) = V_{22}(12) = v_{MF}(12)
\end{equation}
The STA is of the same form as eqs.~(\ref{eq:MBCA-generalForm})--(\ref{eq:L1}), with (\ref{eq:allV}) replaced by
$$V_{12}(12)=V_{21}(12)=V_{22}(12)=u_r(12)$$ 
where $u_r(12)$ is the repulsive part of the potential $u(12)$.

Any other binary collision approximation, as discussed above, leads to results of the form of 
eqs.~(\ref{eq:MBCA-generalForm})--(\ref{eq:L1}),
with the  potentials $V_{12}$, $V_{21}$, and $V_{22}$ being the functions used in the approximation. 
See Table~\ref{table:listApprox} for a list of various approximations we have considered.

\begin{table}
\begin{center}\begin{tabular}{l|cccc}
Approximation & $V_{12}$ & $V_{21}$ & $V_{22}$ & $\hat M_{\hat k\hat
k}(\mathbf k,0)$ \\ \hline
STA  & $\ u_r(r)\ $ & $\ u_r(r)\ $ & $\ u_r(r)\ $ & -332 \\ \hline
BCA  & $u(r)$   & $v_{MF}(r)$  &  $v_{MF}(r)$ &  -281 \\
K-BCA  & $u(r)$   & $v_{MF}(r)$  &  $v_{MF}(r)$ & -281 \\ \hline
MGBE & $v_{MF}(r)$ & $v_{MF}(r)$  &  $v_{MF}(r)$  &  -289 \\
K-MGBE & $v_{MF}(r)$ & $v_{MF}(r)$  &  $v_{MF}(r)$ & -289 \\ \hline
RPMF& $\ v_{rMF}(r)\ $ & $\ v_{rMF}(r)\ $  & $\ v_{rMF}(r)\ $  & -215 \\
K-RPMF & $v_{rMF}(r)$ & $v_{rMF}(r)$  &  $v_{rMF}(r)$  & -215 \\  \hline
hybrid &              &               &                &      \\
MGBE/STA & $v_{MF}(r)$ & $v_{MF}(r)$ & $u_r(r)$ & -289
\end{tabular}\end{center}
\caption{A table of all approximations studied, stating the necessary potential functions that define
a binary collision approximation for the memory function and the value
of the $\hat k\hat k$ matrix element of the memory function at $t=0$. 
The BCA value for $\hat M_{\vec\mu\vec\nu}(\mathbf k,0)$ is exact. Note that $u(r)$
is the bare potential, $u_r(r)$ is the repulsive part of the bare potential, $v_{MF}(r)$ is the potential 
of mean force, and $v_{rMF}(r)$ is the repulsive part of the potential of mean force.
Details about approximations not discussed in this text can be found in Ref.~\onlinecite{jen:thesis}.}
\label{table:listApprox}
\end{table}

Any $K$ approximation, as discussed above, leads to results of the form of 
eqs.~(\ref{eq:MBCA-generalForm})--(\ref{eq:L1}) with eq.~(\ref{eq:fullL}) replaced by
\begin{eqnarray}
\label{eq:Kapprox}
 \lefteqn{\mathcal L_K(12)}\\
   &=& L_0(1) + L_0(2) + \tilde L_1(12) + \tilde L_{BGK}(1) + \tilde L_{BGK}(2) \nonumber
\end{eqnarray}
The symmetry properties of the various approximations are discussed in Appendix~\ref{appendix:sym}.

\section{Calculations}
\subsection{\label{subsct:basisFuncExpan} Basis function expansion of the kinetic equation}
Using the expressions for the $Q$ vertices and $\chi_{BCA}$ response function, the expression in
eq.~(\ref{eq:MBCA-generalForm}) leads to
\begin{eqnarray*}
\lefteqn{M^{(s)}_{BCA}(1t;1't') = \int d2\,d3\,d4\,d5\, \frac{M_m(4)}{M_m(1')}\rho(5)M_m(5)g(45)}\nonumber\\
&&\times\left(\nabla_{\mathbf R}V_{12}(13)\cdot\nabla_{\mathbf P}\delta(12)\right)
\cdot\left(\nabla_{\mathbf R}V_{21}(45)\cdot\nabla_{\mathbf P}\delta(41')\right)\nonumber\\
&&\times\, e^{i\mathcal L(23)(t-t')}\delta(24)\delta(35)\Theta(t-t')
\end{eqnarray*}

It is convenient to take the Fourier transform of $C^{(s)}$ and $M^{(s)}$ with regard to the difference in
their position arguments and to represent their momentum dependence in terms of a set of Hermite polynomial
basis functions. The details of the Hermite polynomial expansion are discussed in
Ref.~\onlinecite{jen:thesis}. From eq.~(\ref{eq:MBCA-generalForm}), we can obtain the following expression
for the Fourier transformed matrix elements of $M^{(s)}$
\begin{eqnarray}
\label{eq:Mmunu}
 \lefteqn{\hat M^{(s)}_{\vec\mu\vec\nu,BCA}(\mathbf k,t) = -\int d\mathbf P_1\,d\mathbf R_2\,d\mathbf P_2 \,
    g(|\mathbf R_1-\mathbf R_2|)} \nonumber\\*
&\times& M_m(\mathbf P_1)\rho M_m(\mathbf P_2)
(\nabla V_{Q_{21}}(|\mathbf R_1-\mathbf R_2|)\cdot \nabla h_{\vec\nu}(\mathbf P_1))\nonumber\\*
&\times& e^{-i\mathbf k\cdot\mathbf R_a} (\nabla V_{Q_{12}}(|\mathbf R_a-\mathbf R_b|)\cdot
    \nabla h_{\vec\mu}(\mathbf P_a))
\end{eqnarray}
The $\vec\mu$ and $\vec\nu$ are labels for the Hermite polynomial basis functions. Each is an ordered
triplet of nonnegative integers. (In the following, we shall use `$\vec 0$' as an abbreviation for the
triplet `000' and `$\hat k$' as an abbreviation for `001'. Note that $ M^{(s)}_{\vec\mu\vec\nu}=0$ if either
$\vec\mu=0$ or $\vec\nu=0$. $M^{(s)}_{\hat k\hat k}$ is the single most important matrix element of the 
memory function. In the limit of small wave vector and/or short time, it is the only matrix element that 
contributes to the correlation function, and in general it is the matrix element that most strongly affects 
the time dependence of the correlation function.) In eq.~(\ref{eq:Mmunu}), $\mathbf R_1$ is defined to be the 
origin and the phase points $\mathbf R_1\mathbf P_1\mathbf R_2\mathbf P_2$ are the initial conditions for 
two particles that evolve forward in time according to Hamilton's equations of motion with $V_{22}$ as the 
interatomic potential to the final phase points $\mathbf R_a\mathbf P_a\mathbf R_b\mathbf P_b$ at time $t$.
The matrix elements of the memory function are calculated using the two particle trajectory calculation method
of Ranganathan and Andersen.\cite{andersen:dkt5} See Appendix~\ref{appendix:numCalc} 
for details and for a discussion of the error analysis.

The correlation functions of interest in this work are the self correlation functions: the incoherent
intermediate scattering function (IISF)
$$ \hat F_s(\mathbf k,t) = \hat C^{(s)}_{\vec0\vec0}(\mathbf k,t) 
    = \frac{1}{N} \left\langle\sum_{i=1}^N e^{-i\mathbf k\cdot(\mathbf r_i(t)-\mathbf r_i(0))} \right\rangle$$
and the self longitudinal current correlation function (SLCC)
\begin{eqnarray*}
 \hat J_{ls}(k,t) &=& \hat C^{(s)}_{\hat k\hat k}(k_z,t) \\ 
  &=& \frac{1}{N} \left\langle\sum_{i=1}^N \frac{p_{iz}(t)}{m}\frac{p_{iz}(0)}{m} 
                e^{-ik_z(r_{iz}(t)-r_{iz}(0))}\right\rangle
\end{eqnarray*}
In these equations it is to be understood that the wave vector $\mathbf k$ is in the $z$-direction.
The IISF and SLCC functions are related in the following way
$$ \hat J_{ls}(\mathbf k,t) = 
         -\frac{1}{|\mathbf k|^2}\frac{\partial ^2}{\partial t^2}\hat F_s(\mathbf k,t) $$
In the graphs below, the SLCC is normalized so that its value at unity is $t=0$.
We focus on the SLCC because we have found its qualitative features to be sensitive to errors in binary
collision approximations. For a discussion of additional correlation functions, see 
Ref.~\onlinecite{jen:thesis}. 

For some of the discussion we will need two related functions. The first is the total correlation function
$$ C(11',t) = \langle \delta f(1,t)\delta f(1',0)\rangle$$
where
$$ f(1,t)=\sum_{i=1}^N \delta(\mathbf R_1-\mathbf r_i(t))\delta(\mathbf P_1-\mathbf p_i(t))$$
is the total particle density and
$$ \delta f(1,t) \equiv f(1,t)-\langle f(1,t)\rangle$$
is a density fluctuation. The second function is the Fourier transform of $C(11',t)$, the coherent intermediate
scattering function (ISF)
$$ \hat F(\mathbf k,t) = \hat C_{\vec0\vec0}(\mathbf k,t) = 
  \frac{1}{N}\left\langle \sum_{i,j=1}^N e^{-i\mathbf k\cdot(\mathbf r_i(t)-\mathbf r_j(0))} \right\rangle $$

\subsection{Solution of the kinetic equations} 
Use of the Hermite polynomial basis set for the momentum arguments converts the kinetic equation for 
$\hat C^{(s)}({\bf k},t)$ into an infinite dimensional matrix equation for the matrix elements of 
$\hat C^{(s)}$. We solve this infinite dimensional equation numerically using the method of Ranganathan and 
Andersen.\cite{andersen:dkt5} In this method, the equations for successively larger finite sets of 
coupled equations are solved. Each of these finite sets is chosen such that its solution agrees with the 
exact solution of the infinite set up to order $t^n$ for small $t$, and successive finite sets have 
increasing values of $n$. When the successive solutions no longer change with increasing $n$, the last 
result is regarded as the solution to the infinite set. A detailed discussion of the method can be found in 
Ref.~\onlinecite{mr:thesis}.

The IISF, which corresponds to $\hat C^{(s)}_{\vec 0\vec 0}(\mathbf k,t)$, is converged with the 
$\mathcal O(t^9)$ approximation. The SLCC, which corresponds to 
$\hat C^{(s)}_{\hat k\hat k}(\mathbf k,t)$, is converged with the $\mathcal O(t^7)$ approximation. We present 
only these solutions in the results that follow. Moreover, we concentrate on the results for the SLCC since 
none of the approximations are able to consistently reproduce all of the features of this function, especially
the temperature dependence of the negative region for the function.
Since the SLCC is proportional the second derivative of the IISF, the accuracy of an 
approximation for the SLCC function is consistent with its accuracy for the IISF.

\subsection{Molecular dynamics calculations}
The correlation functions calculated from the binary collision approximations for the memory function will 
be compared to results generated from molecular dynamics simulations of a 500-particle, one component 
Lennard-Jones liquid
at equilibrium.\cite{tomMDI,tomMDII,stein} The states of interest are reduced density $\rho=0.85$
(approximately the triple point density) at reduced temperatures $T = 0.723,\, 1.554,\, 3.000$
(the triple point temperature is approximately $0.723$). Results for temperatures between 0.723 and 1.554 
at $\rho=0.85$, and high temperature results at $\rho=0.75$ can be found in Ref.~\onlinecite{jen:thesis}.
The usual reduced units for the Lennard-Jones fluid are used.\cite{AllenTildesley}
The correlation function results from molecular dynamics simulation and the binary collision approximations
have small statistical error whose values are generally smaller than the differences between the simulation 
data and the theoretical curves. The details of the molecular dynamics simulations can be found in 
Refs.~\onlinecite{tomMDI} and \onlinecite{tomMDII}.

\section{Results} 
\subsection{BCA}
Although the BCA approximation gives exactly the correct $t=0$ value of the memory function, this approximation 
leads to correlation functions that oscillate rapidly and with sharply increasing magnitude as $t\rightarrow\infty$. 
Although the shape of the BCA memory function
looks reasonable, there is insufficient area under the curve for the solution to the kinetic equation to be 
well-behaved. We suspect that this is related to its failure to satisfy the symmetry property discussed in 
Appendix~\ref{appendix:sym}. (Recall the form of this approximation, 
eqs.~(\ref{eq:MBCA-generalForm})--(\ref{eq:L1}).)

\subsection{STA, MGBE, and K-MGBE}

\subsubsection{STA}
Overall, for the high densities studied, the STA approximation does reasonably well at predicting correlation 
functions at all temperatures and wave vectors, with increasing quantitative accuracy with high temperature 
and large wave vector. The short time predictions of the STA are reasonably accurate, as is the overall 
longer time behavior. The STA approximation is able to reproduce qualitative trends of the various 
correlation functions with changes in wave vector. The trends with changes in temperature are also reproduced, 
except for the SLCC curve where the
shifting and increasing magnitude of the negative dip is not well reproduced. In fact, the STA approximation
consistently fails to predict the temperature dependence of the negative dips in the SLCC curve.
A complete discussion of the results of the STA approximation can be found in 
Refs.~\onlinecite{andersen:dkt5} and \onlinecite{jen:thesis}.

\subsubsection{MGBE}
\begin{figure}
\epsfig{figure=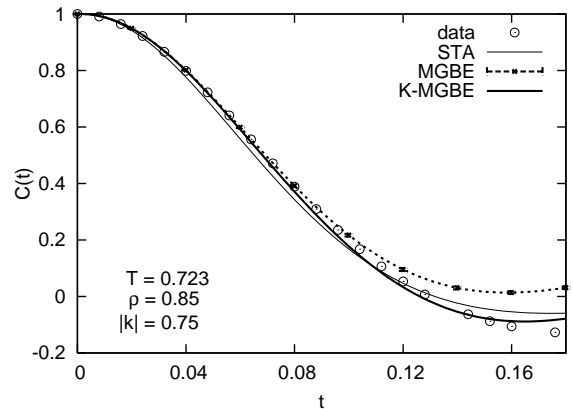,width=8cm}
\caption{Normalized SLCC function. Although the MGBE and K-BCA approximations do poorly at representing
the simulation result for longer times, for times $t<0.1$ we see that these are the most quantitatively
accurate approximations. The statistical error bars for the MGBE correlation function are shown and are 
barely discernable on this scale. 
All of the theoretical correlation functions have similarly small statistical errors.}
\label{fig:st-SLCC-lowTk}
\end{figure}

The MGBE approximation is more accurate at short times than the STA, as shown in Fig.~\ref{fig:st-SLCC-lowTk},
and is in essentially perfect agreement with the simulation data for times less than about 
0.1. This statement holds for the self correlation functions at all 
temperatures and wave vectors studied.  The short time part of the SLCC function is largely determined by 
the value of the $\hat k\hat k$ matrix element of the memory function at $t=0$. The MGBE does not have 
exactly the correct zero time values of the memory function matrix elements but it is more accurate than 
the STA, and this accounts for its accuracy at short times.   The MGBE result begins to become inaccurate at
$t\approx0.1$, rising significantly above the simulation data.

\begin{figure*}
\epsfig{figure=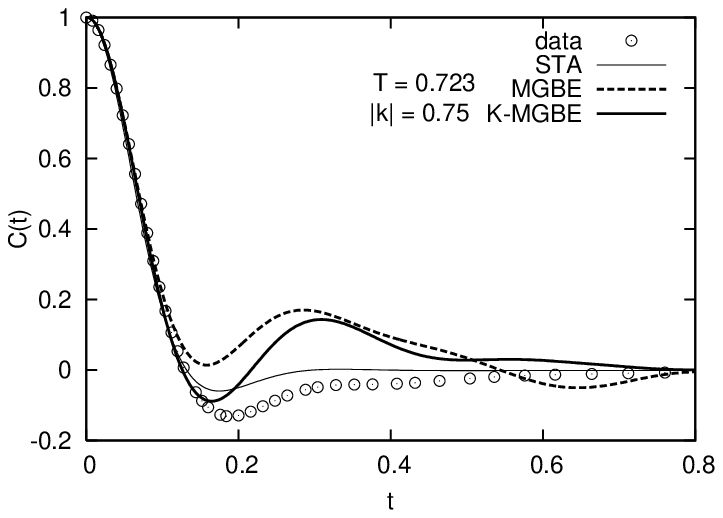,width=8cm}
\epsfig{figure=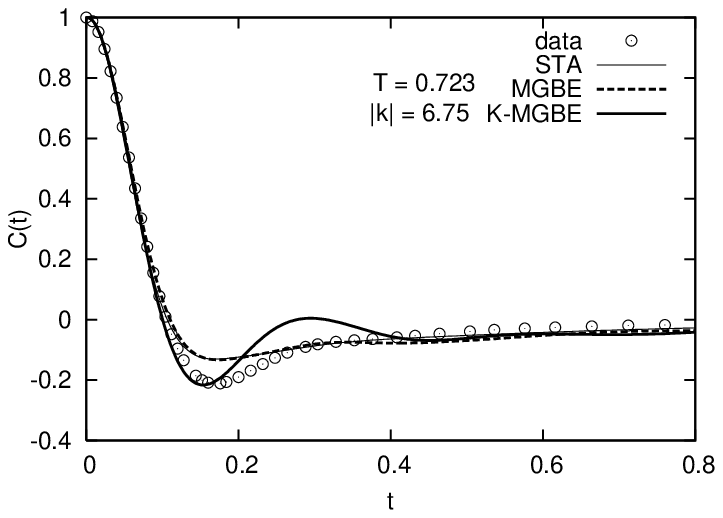,width=8cm}
\caption{Normalized SLCC function for the STA, MGBE and K-MGBE approximations for $T=0.723$ at small and 
large wave vector. The STA and MGBE curves lie on top of each other at large wave vector. 
High temperature results are shown in Fig.~\ref{fig:SLCC-lt-compare-highT}.}
\label{fig:SLCC-lt-compare-lowT}
\end{figure*}

\begin{figure*}
\epsfig{figure=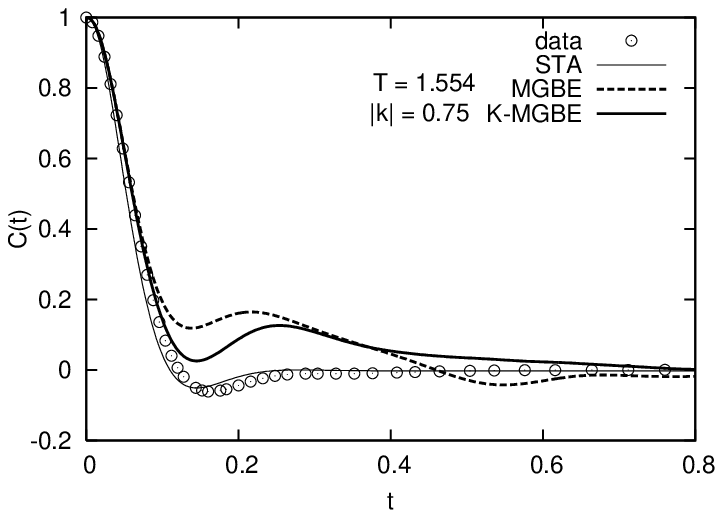,width=8cm}
\epsfig{figure=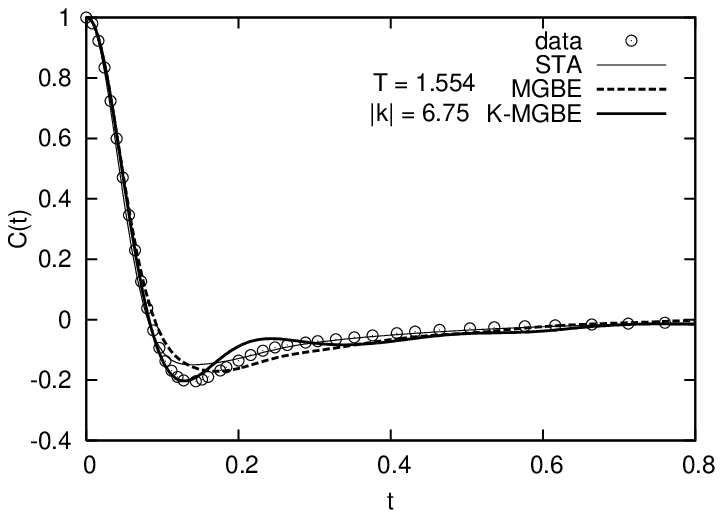,width=8cm}
\caption{Normalized SLCC results for the STA, MGBE, and K-MGBE approximations for $T=1.554$ at small and
large wave vector. See Fig.~\ref{fig:SLCC-lt-compare-lowT} for low temperature results. }
\label{fig:SLCC-lt-compare-highT}
\end{figure*}

For longer times, $t\ge0.1$, the MGBE gives very inaccurate results. 
See Figs.~\ref{fig:SLCC-lt-compare-lowT} and \ref{fig:SLCC-lt-compare-highT}.  For large 
wave vectors at both high and low temperatures, the MGBE behaves reasonably, with a single minimum in the 
SLCC at approximately the right time, but an overall quantitative disagreement with the data.  For small wave 
vectors at all temperatures, it predicts oscillation in the SLCC for times of about 0.2-0.3, a positive 
maximum in the function at $t\approx0.3$, and a very slow decay of the correlation function to zero from 
above, in qualitative disagreement with the simulation results.  The latter have a single minimum, followed
by an approach to zero from below on a shorter time scale than the MGBE result approaches zero.

The origin of this unusual behavior, especially the oscillations, is easily understood from the nature of 
the MGBE.  The memory function in the MGBE is a time correlation function for derivatives of the potential 
of mean force in a two particle system for which the two particles interact according to the potential of mean 
force of the fluid and move according to Hamiltonian dynamics. See eqs.~(\ref{eq:MBCA-generalForm}),
(\ref{eq:MGBE-allV}), (\ref{eq:Mmunu}) and Appendix~\ref{appendix:numCalc}.
This idea, which follows from the explicit formula eq.~(\ref{eq:Mmunu}), 
is the basis for the algorithm used to evaluate the matrix 
elements of the memory function.  The potential of mean force at high density is oscillatory and extends 
to distances of several atomic diameters.  Thus, collisions that are relevant for the MGBE are of long duration 
and the correlation function that gives the memory function is nonzero for a very long time.  

\begin{figure}
\epsfig{figure=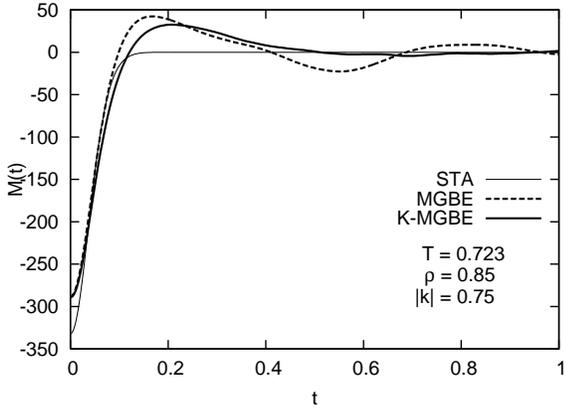,width=8cm}
\caption{Self part of the $\hat k\hat k$ matrix element of the memory function.
The STA approximation generates a non-oscillatory memory function while both MGBE and K-MGBE approximations 
yield oscillatory memory functions. These oscillations are the cause of the oscillations in the 
MGBE and K-MGBE results for the SLCC.}
\label{fig:MF-shortTime-comp}
\end{figure}

Figure~\ref{fig:MF-shortTime-comp} 
shows the time dependence of the principal matrix element of the memory function for the MGBE 
approximation for a low temperature state. There is a pronounced oscillation that persists for times 
longer than the time scale for the true correlation function to decay.  For comparison, the STA memory
function decays to zero very quickly, within times of the order of 0.1.  As we shall discuss below in the 
next section, the large positive feature in the memory function and the subsequent oscillation are not  
present in the actual memory function.

These results show that the MGBE approximation, although quantitatively correct at short times, 
is qualitatively incorrect at intermediate and long times. 

\subsubsection{K-MGBE}  
The collisions described above that are the basis of the MGBE approximation, namely Hamiltonian dynamics 
of a pair of particles whose interaction potential is equal to the potential of mean force of the liquid, 
seem inappropriate for determining the dynamics of a dense liquid.  The particles surrounding a pair of 
particles should do more to determine the dynamics of the particle pair than merely to renormalize 
the bare potential into the potential of mean force. In Mazenko's theory, this would be taken into account 
by including the quantity called $\Gamma$, which represents renormalized dynamical interactions of the pair.  
In the cluster theory, this would be taken into account by including the coupling of the pair to fluctuations 
of the density of third particles.

The K-MGBE approximation discussed above provides a tractable way of including these interactions, provided 
they are regarded as independent uncorrelated repulsive collisions of the two particles with third particles, 
approximated using the BGK model memory function.  This approximation sums, in an approximate way, a class of 
diagrams in the exact diagrammatic theory.  We assume that these collisions involve the short range repulsive 
part of the interatomic potential, so we pick the parameter $\nu$ so that the BGK memory function has a 
strength that is related to that of the STA memory function. Since the most important matrix element of the 
memory function is the $\hat k\hat k$ matrix element, we choose $\nu$ so that this matrix element for the BGK 
model corresponds to the actual matrix element for the STA.  More precisely, we choose $\nu$ such that 
\begin{equation}
\label{eq:nu}
\nu = \int_0^\infty dt'\,\hat M^{(s)}_{\hat k\hat k,STA}(\mathbf0,t')  
\end{equation}
While this is a crude approximation overall, it was devised to give some idea of the effects of third particles 
on the dynamics of the pair, in the hope of yielding an improved version of the MGBE.

Figure~\ref{fig:MF-shortTime-comp} 
shows that the K-MGBE memory function has less pronounced oscillations and a more rapid decay to 
zero than the MGBE memory function.  Figures~\ref{fig:SLCC-lt-compare-lowT} and \ref{fig:SLCC-lt-compare-highT}
show the SLCC correlation functions given by the K-MGBE 
approximation. Introducing the BGK collisions extends the initial time interval over which the approximation 
is valid, for low and high temperatures and for small and large wave vector. The intermediate time behavior 
of the K-MBGE is unsatisfactory, however, with a peak in the correlation function at times of about 0.3 
appearing under all conditions studied.  The long time behavior of the K-MGBE is at best a slight improvement 
over that of the MGBE.

\subsection{Additional approximations}
In an attempt to find improved binary collision approximations, we investigated a number of other 
possibilities.

\paragraph{RPMF approximation.}

The STA memory function, which uses the repulsive part of the potential in each of the vertices, has the 
virtue of decaying to zero rapidly without oscillations at long time that produce unphysical behavior of the 
correlation function at long times.  The rapid decay is a result of the fact that the two particle dynamics 
in the calculation of the memory function involves purely repulsive interactions (and hence brief collisions) 
and the range of the repulsive potential is so short.

An approximation, which we denote RPMF, based on using the repulsive part of the potential of mean force in 
each vertex, was constructed and studied.  While the memory function  generated by this approximation has 
the feature of rapid decay, as discussed for the STA, its magnitude and its integrated area are smaller than 
that of the STA and it generates an SLCC that falls from its initial value significantly more slowly that the 
STA result.  This basic flaw in the RPMF approximation is evident in the zero time value of the memory 
function (see Table~\ref{table:listApprox}).  

\paragraph{K-RPMF approximation.}  
The K-RPMF approximation adds BGK collisions with third atoms to the RPMF approximation.  This memory 
function decays more slowly than that of the RPMF, but it has the same initial value as the RPMF, and 
overall  is not a significant improvement over the RPMF.

\paragraph{A hybrid of the MGBE and STA approximations.}  
We attempted to construct approximations that have accurate values of the memory function at zero time but 
that were based on relatively brief collisions.  In one such approximation, we used the potential of mean 
force in the $Q_{12}^{(c,1)}$ vertex and the $Q_{21}^{(c,1)}$ vertex (as in the MGBE approximation) but the 
repulsive part of the potential in the $Q_{22}^{(c,1)}$ vertex (as in the STA).  We also considered the 
approximation that supplemented this approximation with BGK collisions with third particles. This hybrid 
is not a significant improvement over the MGBE or STA approximations 
and had some artifacts for small wave vector that are inconsistent with the simulation results.

\paragraph{Other approximations.}  
A variety of other unsuccessful approximation were also studied.  See Ref.~\onlinecite{jen:thesis} for details.

\section{Extraction of the memory function}
Thus far, we have tested approximate kinetic theories by calculating their memory functions, solving the 
memory function equation for the related correlation functions, and comparing the predicted correlation 
functions with computer simulation results. These comparisons give only an indirect test of the accuracy of
the approximate memory function approximations.

In this section we consider a different set of questions.  Given a correlation function calculated from
simulation data, 
what is the memory function that corresponds to it? What are the properties of the `exact' memory function?  
How do the approximate memory functions compare with the `exact' memory function?

The numerical technique we use for extracting the memory function from the correlation function data 
employs a Fourier series representation of the memory function as a function of time.  The Fourier 
coefficients in the series were adjusted to minimize the differences between the SLCC calculated from the 
memory function and the SLCC obtained from computer simulation.  For small wave vector, the $\hat k\hat k$ 
matrix element of the memory function plays the dominant role in determining the SLCC 
function.\cite{jen:thesis}  
The reason for this is that the coupling of the correlation function to other elements of the memory function 
involves positive powers of the wave vector.  Thus we need to construct a Fourier representation of only this 
matrix element of the memory function and solve for the Fourier coefficients of just one function rather 
than many.  The coefficients are found by a minimization of the sum of the squares of the deviations of the 
calculated correlation function and the simulated function.  See Ref.~\onlinecite{jen:thesis} for details.

\begin{figure}
\epsfig{figure=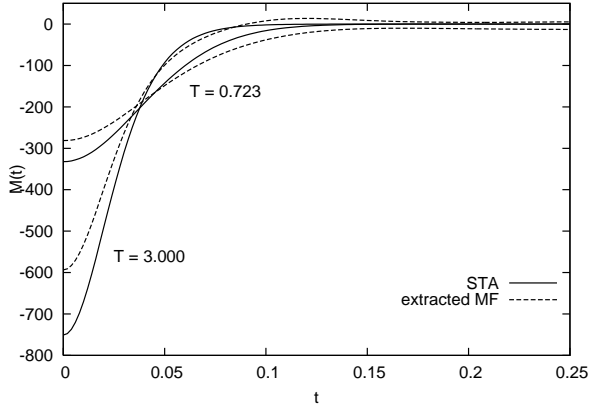,width=8cm}
\caption{Extraction results (dashed lines) for the $\hat k\hat k$ matrix element of the self memory 
function compa STA curves (solid lines) at $\rho=0.85$ and $|\mathbf k|=0.75$. 
The top set of lines at $t=0$ corresponds to $T=0$. The lower set at $t=0$ corresponds to $T=3.000$.}
\label{fig:extractMF-results}
\end{figure}

The extracted $\hat M^{(s)}_{\hat k\hat k}({\bf k},t)$ for small wave vector is shown in 
Fig.~\ref{fig:extractMF-results} for the 
lowest and highest temperatures studied, where they are compared with the STA predictions for the same 
function.  The extracted memory function at low temperature has none of the oscillations and high maxima 
that are predicted by the MGBE and K-MGBE approximations.  
Compare Fig.~\ref{fig:MF-shortTime-comp}. The STA results are qualitatively
similar to  the extracted results, but there are significant differences.  1.~The magnitude at zero time 
is larger for the STA than for the extracted memory function. 
2.~The STA curves fall rapidly to zero, while the extracted curves 
decay to zero more slowly, from above at the higher temperature, and from below at lower temperatures.  

\begin{figure}
\epsfig{figure=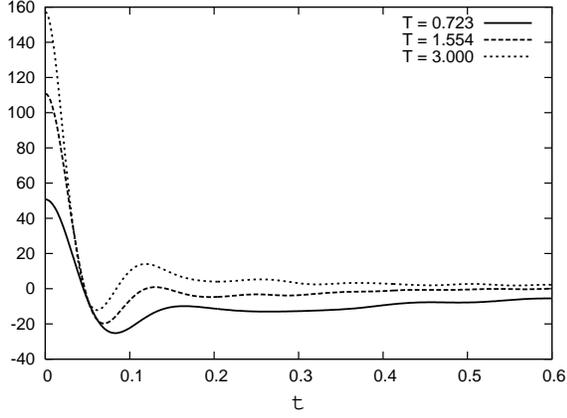,width=8cm}
\caption{The difference between the extracted and STA memory functions, 
$\hat M_{kk,STA}(0.75,t)-\hat M_{kk,ext}(0.75,t)$, for three different temperatures.}
\label{fig:extractMF-diff}
\end{figure}

\begin{figure}
\epsfig{figure=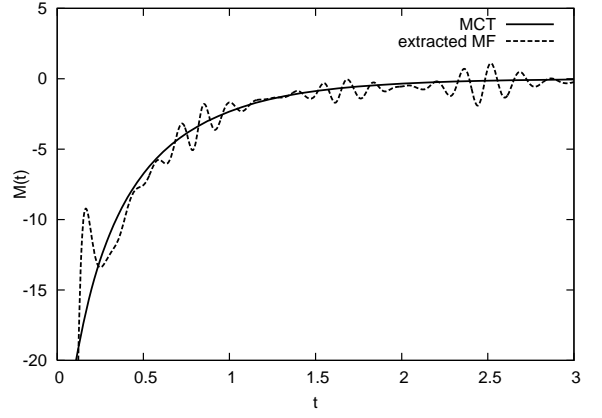,width=8cm}
\caption{The longer time part of the memory function predicted by mode coupling theory is compared to the 
memory function numerically extracted from the simulation data for the SLCC at $T=0.723$ 
and $|\mathbf k|=0.75$.}
\label{fig:mctTest}
\end{figure}

The differences between the extracted memory function and the STA approximation is plotted for high, medium, 
and low temperatures in Fig.~\ref{fig:extractMF-diff}. At the highest 
temperature, the memory function decays to zero from above.  This is presumably a manifestation of the 
hydrodynamic vortex effect that is seen for hard spheres.\cite{alder} The long time negative part 
of the memory function at lower temperatures is presumably a manifestation of the cage effect, and is related 
to the long slow decay of the simulation correlation function at low temperature and small wave vector 
(see Fig.~\ref{fig:SLCC-lt-compare-lowT}).  
The very long time part of the memory function is shown in Fig.~\ref{fig:mctTest}.

The actual memory function at low temperatures clearly has a rapidly decaying part, falling to 2\% 
of its initial value at $t\approx0.5$, and a  more slowly decaying part that falls to zero (within the 
statistical error) at about $t=1.5$.  None of the binary collision approximations we have studied is 
qualitatively consistent with the true long time behavior of the memory function.   
Moreover, none of them provides a quantitative description of the initial rapid decay.  

\section{Mode coupling description of the long time part of the memory function}
The mode coupling theory of G\"{o}tze\cite{gotze} can be used to calculate, among other things, a 
correlation function denoted $\phi^{(s)}_q(t)$ that is proportional to 
$\hat C^{(s)}_{\vec0\vec0}(\mathbf k,t)$, the Fourier transform of a specific matrix element of 
$C^{(s)}(11',t)$. The theory is based on the Mori-Zwanzig projection operator method,\cite{mori:I} 
which relates this 
function to its second order memory function, denoted $M^{(s)}_q(t)$. The assumptions of the theory lead 
to a mode coupling relationship that expresses the long time part of $M^{(s)}_q(t)$ as a bilinear 
functional of $\phi^{(s)}_q(t)$ and $\phi_q(t)$, where the latter is proportional to 
$\hat C_{\vec0\vec0}(\mathbf q,t)$. See eq.~(3.34) of Ref.~\onlinecite{gotze}.

The kinetic theory presented here relates $\hat C^{(s)}_{\vec0\vec0}(\mathbf k,t)$ to an infinite 
dimensional matrix of memory functions $\hat M^{(s)}_{\vec\mu\vec\nu}(\mathbf k,t)$. For small 
wave vector $\mathbf k$, $\hat C^{(s)}_{\vec0\vec0}(\mathbf k,t)$ is coupled primarily to 
$\hat M^{(s)}_{\hat k\hat k}(\mathbf k,t)$, and in the limit as this wave vector goes to zero, this is the 
only element that affects the value of $\hat C^{(s)}_{\vec0\vec0}(\mathbf k,t)$. Our numerical work 
confirms that $|\mathbf k|=0.75$ is small enough for this to be the case. It is straightforward to show 
that when this is the case, the relationship between $\hat M^{(s)}_{\hat k\hat k}(\mathbf k,t)$ 
and $\hat C^{(s)}_{\vec0\vec0}(\mathbf k,t)$ is precisely of the form generated by the Mori-Zwanzig 
theory. In other words, for small wave vector, the single matrix element 
$\hat M^{(s)}_{\hat k\hat k}(\mathbf k,t)$  is the Mori-Zwanzig second order memory function of 
$\hat C^{(s)}_{\vec0\vec0}(\mathbf k,t)$.

When the mode coupling relationship mentioned above is expressed in terms of the present kinetic theory, 
the result is of the following form
$$ \hat M^{(s)}_{\hat k\hat k}(\mathbf k,t) 
   = \int d\mathbf q\,A(\mathbf k,\mathbf q)
     \hat C_{\vec0\vec0}(\mathbf q,t)\hat C^{(s)}_{\vec0\vec0}(\mathbf k-\mathbf q,t) $$
where $A(\mathbf k,\mathbf q)$ is a nonnegative function. The mode coupling theory asserts that this 
holds for times longer than the duration of brief binary collisions, which in the present theory 
corresponds roughly to the time at which the STA memory function has decayed to zero.

Applying this result to $|\mathbf k|=0$, we find
\begin{equation}
 \hat M^{(s)}_{\hat k\hat k}(\mathbf 0,t) = 
  \int d\mathbf q\,A(\mathbf 0,\mathbf q)
           \hat C_{\vec0\vec0}(\mathbf q,t)\hat C^{(s)}_{\vec0\vec0}(\mathbf q,t)
\end{equation}
Each of the two time dependent functions on the right depends on the magnitude, but not the direction, 
of its wave vector argument. The function $A(\mathbf 0,\mathbf q)$ is zero if $|\mathbf q|=0$, small 
near $|\mathbf q|=0$, and highly peaked near $|\mathbf q|=q_{max}\approx 2\pi/\sigma$. Assuming that the 
integral above is dominated by the range of wave vectors whose magnitude is near this value, we find
\begin{equation}
\label{eq:mctPrediction}
  \hat M^{(s)}_{\hat k\hat k}(0,t) \approx 
     (constant) \times \hat C_{\vec0\vec0}(q_{max},t)\hat C^{(s)}_{\vec 0\vec0}(q_{max},t)
\end{equation}
where the constant is positive. Since $\hat M^{(s)}_{\hat k\hat k}(\mathbf k,t)$ is an even differentiable 
function  of the wave vector argument, it is very slowly varying near $|\mathbf k|=0$, and we will apply
this result for $|\mathbf k|=0.75$, the smallest nonzero wave vector studied in this work.

We can test this relationship by comparing the time dependence of the extracted 
$\hat M^{(s)}_{\hat k\hat k}(0.75,t)$ obtained from the
simulation data with the time dependence of the right side of eq.~(\ref{eq:mctPrediction}), which contains 
functions that are routinely obtained from simulation data. Figure~\ref{fig:mctTest} shows the extracted 
memory function $\hat M^{(s)}_{\hat k\hat k}(0.75,t)$ for longer times ($t>0.1$) for the lowest temperature 
studied. At that time, the STA memory function has decayed almost completely to zero. The extracted memory 
function has oscillations that are an artifact of the extraction procedure (note the vertical scale compared 
with the graph of Fig.~\ref{fig:extractMF-results}). Figure~\ref{fig:mctTest} also shows the right side of 
eq.~(\ref{eq:mctPrediction}), with the constant chosen so that the results matches the value of the extracted 
memory function for $t\approx 1$. With this choice of the constant, the time dependence of the right side 
accounts remarkably well for the time dependence of the memory function matrix element.

Note that this test was only for the smallest wave vector at the lowest temperature studied. For high 
temperatures, eq.~(\ref{eq:mctPrediction}) is qualitatively incorrect ({\it i.e.,} the predicted sign 
of the memory function is incorrect). This is presumably because the behavior is dominated by a hydrodynamic 
vortex effect,\cite{alder} rather than the cage effect described by mode coupling theory. 
For higher wave vectors
at low temperatures, the test cannot be performed, since the simple relationship between 
$\hat M^{(s)}_{\hat k\hat k}(\mathbf k,t)$ and the second order memory function of the Mori-Zwanzig theory 
does not hold. Note, also, that mode coupling theory allows, in principle, the calculation of the constant 
of proportionality, but we have not done this. Nevertheless, the agreement in Fig.~\ref{fig:mctTest} is 
striking, suggesting that mode coupling theory accounts quantitatively for the time dependence of the memory 
function for long times ({\it i.e.,} for times longer than the time of a brief binary collision) for the 
lowest temperature studied, which corresponds to the liquid state near the triple point.

\section{Summary and Conclusions}
\setcounter{paragraph}{0}
\paragraph{The accuracy of various approximate theories.}We have investigated various binary collision 
approximations for the memory function of the correlation function of density fluctuations in a dense Lennard-Jones 
liquid at equilibrium and compared their predictions with computer simulation results for the self 
correlation functions, especially the self longitudinal current correlation function (SLCC).

The BCA of Ranganathan and Andersen is exact at $t=0$, but it leads to divergent results at long times, 
presumably because of its failure to satisfy important symmetry properties.

The moderate density approximation of the generalized Boltzmann-Enskog memory function (MGBE) of Mazenko and 
Yip\cite{mazenko77} is accurate at very short times. However, its intermediate time behavior is very different 
from the correct results, and its predicted correlation functions display unphysical oscillations at intermediate 
times.

The STA of Ranganathan and Andersen\cite{andersen:dkt4,andersen:dkt5} is not exact at $t=0$. It describes the 
initial drop of the SLCC to zero reasonably accurately, but it does not accurately describe the shape and depth 
of the minimum in that function and its temperature dependence. At high temperatures, it gives accurate values 
of the transport coefficients.\cite{andersen:dkt5}

The K-MGBE approximation, which introduces uncorrelated brief collisions of the colliding pair with third 
particles, is more accurate than the MGBE at short times but its intermediate time behavior is still 
very inaccurate.

No other binary collision approximation discussed in this paper is a systematic improvement over the STA.

\paragraph{The behavior of the extracted memory function.}  
For small wave vector, the most important matrix element of the memory function can be extracted from the 
simulation results for the SLCC.  

At short time, the extracted memory function drops rapidly to values close to zero, in a way similar to the STA 
result, and then has a very small, long-lived, nonoscillatory tail that is positive at high temperatures and 
negative at low temperatures.  At low temperatures, the negative tail is important for determining the shape of 
the minimum in the SLCC.  This tail is probably the result of physical effects that are not included in any 
binary collision approximation.  The time dependence of the long time tail of the memory function is in striking 
agreement with what is predicted by mode coupling theory, but this is only a limited confirmation of that theory
since the extracted memory function can be obtained only at small wave vector.

\paragraph{Conclusions and speculations.}  
The negative feature of the SLCC for the dense Lennard-Jones fluid indicates that the velocity of a particle is 
anticorrelated with its velocity at times that are more than about 0.1 time units earlier.  An important part of 
the basis for this feature can be described by binary collision theories that take into account the finite 
duration of very repulsive collisions.  For hard spheres, the short time feature of the memory function is a Dirac 
delta function, which by itself would generate monotonic decay of the SLCC from above.  In the STA, the short 
time feature is spread over a range of times of the order of the duration of a repulsive collision, and the 
resulting correlation function has a negative dip. However, the increase of the depth of the negative feature as 
the temperature is lowered is caused not by this short time feature but rather by a small, negative, temperature 
dependent long time tail in the memory function.  This is presumably a manifestation of what is commonly regarded 
as caging of atoms at low temperatures and cannot be described by binary collision approximations.  
Instead, mode coupling theory or something else that goes beyond binary collision approximations is needed to 
describe this aspect of the negative feature.

This conclusion is not at all surprising.  Theories that  use mode coupling commonly include a binary collision 
term in the memory function but do not discuss in detail the nature of that term.  It is clear from the present 
work that the short time part of the memory function has behavior associated with brief binary {\em repulsive} 
collisions, such as those described by the STA.  Collisions that include attractive as well as repulsive 
interactions, such as those of the MGBE, have a much longer duration, and theories that include them have 
memory functions that decay to zero much too slowly to provide a good first approximation for the correlation 
functions at intermediate times.

We have not been able to derive any binary collision theory that describes brief binary repulsive collisions and 
that has exactly the correct value of the memory function at $t=0$.  There is a straightforward expression for 
the zero time value of the memory function\cite{note9}
in terms of static correlation functions, and this expression involves the entire potential, 
the entire potential of mean force, and the pair correlation function.  Any graphical approximation for the 
memory function that includes all this at $t=0$ would have more physics in it than binary repulsive collisions. 

This leads us to speculate that the memory function for density fluctuations can be  usefully regarded as a 
sum of at least three parts: a contribution from repulsive binary collisions (the STA or something similar 
to it), another short time part that is related to all the other interactions (but whose nature is not 
understood), and a longer time slowly decaying part that describes caging (of the type predicted by mode 
coupling theory).

\acknowledgments
The authors would like to thank Dr.\ Thomas Young and Richard Stein for providing the molecular dynamics data
used in this work. This work was supported by the National Science Foundation grant CHE-0408786.
One of the authors (J.N.-V.) would like to acknowledge support from a DOE CSGF fellowship.

\appendix
\setcounter{paragraph}{0}

\section{\label{appendix:sym} Symmetry properties of the memory function}
The theory we are discussing describes a system with a time independent Hamiltonian at equilibrium.  
The exact correlation function for the system must satisfy symmetry properties related to stationarity and 
time reversal invariance.  Stationarity implies
\begin{equation}
\label{eq:stationarity}
C^{(s)}({\bf R}_1,{\bf P}_1;{\bf R}_1^\prime,{\bf P}_1^\prime;t) =
    C^{(s)}({\bf R}_1^\prime,{\bf P}_1^\prime;{\bf R}_1,{\bf P}_1;-t)
\end{equation}
A combination of time reversal symmetry and stationarity implies
$$C^{(s)}({\bf R}_1,{\bf P}_1;{\bf R}_1^\prime,{\bf P}_1^\prime;t) =
     C^{(s)}({\bf R}_1^\prime,-{\bf P}_1^\prime;{\bf R}_1,-{\bf P}_1;t)$$
There are similar implications for other correlation functions and for the memory function. The importance 
of related symmetry conditions formulated in the Laplace transform domain have been emphasized by 
Mazenko.\cite{mazenko77}

The exact diagrammatic series satisfies these relationships exactly.  This follows from the fact that the 
series are formally exact, but it can also be proven directly from the topological structure of the diagrams 
and the symmetry properties of the vertices and bonds that appear in the diagrams.  (The diagrammatic theory 
was derived \cite{andersen:dkt1,andersen:dkt2,andersen:dkt3,andersen:dkt4,andersen:dkt5} only for positive 
time arguments.  However, the results can easily be extended to negative time by analytic continuation in 
order to verify that they satisfy eq.~(\ref{eq:stationarity}).)

Graphical approximations are constructed by retaining a subset of diagrams in the exact series and/or by 
making approximations for the vertices. Such graphical approximations may or may not satisfy the symmetry 
requirements, depending on the nature of the subset retained and details of the approximations made.  

We have not been able to formulate a useful set of necessary conditions that a general graphical approximation 
must satisfy in order to be consistent with the symmetry properties. However,  various sets of sufficient 
conditions can be formulated and proven.  In particular, two sets of sufficient conditions for a binary 
collision approximation, of the class discussed above, to satisfy the symmetry property can be stated. 
The sufficient conditions reduce to conditions on the quantities that appear in the vertices.  
The details will be omitted here and we shall merely quote the results.
1.~The MGBE approximation for $M^{(s)}$ satisfies the symmetry conditions exactly.
2.~The BCA approximation for $M^{(s)}$ does not satisfy the known sufficient conditions for symmetry.  
It is likely that it strongly fails to satisfy the symmetry, and this is probably the reason why it is such 
a poor approximation at long times.
3.~The STA approximation for $M^{(s)}$ does not satisfy either set of known sufficient conditions for symmetry 
exactly. However, in the limit in which the repulsive forces become hard sphere forces, it does satisfy 
one set.  Thus for very strong repulsive forces it is at least numerically close to satisfying the symmetry 
conditions. 
4.~The RPMF approximation for $M^{(s)}$ satisfies the symmetry conditions exactly. 
5.~The hybrid approximation discussed above does not satisfy either set of known sufficient conditions for 
symmetry exactly.  
6.~If a binary collision approximation satisfies either of the two known sets of sufficient conditions, 
its $K$ approximation satisfies the symmetry conditions exactly. Thus the K-MGBE and K-RPMF satisfy the 
symmetry properties exactly.

\section{\label{appendix:leftVertex} Two forms of the left vertex in the memory function}
\paragraph{The left vertex in the MGBE and BCA.}
The reasoning that leads to the derivation of the MGBE approximation in the context of the Mazenko theory 
and the BCA in the context of the graphical theory are very similar, but the results are slightly different.  
The nature of the difference is discussed in this appendix.

The graphical theory contains a vertex $Q_{12}(1;1^\prime2^\prime)$ defined as
\begin{eqnarray}
\lefteqn{Q_{12}(1;1^\prime2^\prime)} \nonumber\\
\label{eq:definitionQ12}
  &\equiv&\frac{1}{2!}\int d1^{\prime\prime}d2^{\prime\prime}\,W_{12}(1;1^{\prime\prime}2^{\prime\prime})
    K_2(1^{\prime\prime}2^{\prime\prime};1^\prime2^\prime) \\
\label{eq:exactQ12}
&=&   \left(1+{\cal I}(1^\prime2^\prime)\right)\nabla_{\mathbf R}
        u(12^\prime)\cdot\nabla_{\mathbf P}\delta(11^\prime) 
\end{eqnarray}
(See eq.~(3) and Sec.~III of Ranganathan and Andersen\cite{andersen:dkt4} and Appendix A.4 of 
Andersen.\cite{andersen:dkt2} Here the ${\cal I}$ operator interchanges the two arguments in the subsequent 
expression.) The first equality is a definition. The second equality represents an exact evaluation.  
Although this function has three arguments, they represent only two distinct particles, since the Dirac delta 
function in each term clearly indicates that the left argument corresponds to the same particle as one of the 
right arguments.  So this vertex appears appropriate for a theory that includes binary collisions between 
particles.

An exact evaluation of $W_{12}(1;1^\prime2^\prime)$ gives two terms.  The first term is
\begin{eqnarray*}
\lefteqn{W_{12}(1;1^\prime2^\prime)_{\mathrm{first\,term}}
 =  -(1+{\cal I}({1^{\prime\prime}2^{\prime\prime}}))n(1^{\prime\prime})n(2^{\prime\prime})} \\
    &\times& M_m(1^{\prime\prime})M_m(2^{\prime\prime})g(1^{\prime\prime}2^{\prime\prime})\nabla_{\mathbf P}
        \delta(1^{\prime\prime}1)\cdot\nabla_{\mathbf R}v_{MF}(1^{\prime\prime}2^{\prime\prime})
\end{eqnarray*}
It clearly represents two distinct particles. The second term (not given here), which is very complicated, 
involves static correlations functions for three distinct particles.

An exact evaluation of $W_{12}(1;1^\prime2^\prime)$ gives two terms.  The first term is
\begin{eqnarray*}
\lefteqn{W_{12}(1;1^\prime2^\prime)_{\mathrm{first\,term}}
 = -(1+{\cal I}({1^{\prime\prime}2^{\prime\prime}}))n(1^{\prime\prime})n(2^{\prime\prime})} \\
 &\times&  M_m(1^{\prime\prime})M_m2^{\prime\prime})g(1^{\prime\prime}2^{\prime\prime})
   \nabla_{\mathbf P}\delta(1^{\prime\prime}1)\cdot\nabla_{\mathbf R}v_{MF}(1^{\prime\prime}2^{\prime\prime})
\end{eqnarray*}
This clearly represents two distinct particles.  The second term (not given here), which is very complicated, 
involves static correlations functions for three distinct particles.

An exact evaluation of $K_2(1^{\prime\prime}2^{\prime\prime};1^\prime2^\prime)$ also gives two terms.  
The first term is
\begin{eqnarray}
\lefteqn{K_2(1^{\prime\prime}2^{\prime\prime};1^\prime2^\prime)_{\mathrm{first\,term}}}
\nonumber
\\*
\label{eq:K2firstterm}
 &=&\frac{\left(1+{\cal I}(1^{\prime\prime}2^{\prime\prime})\right)\delta(1^{\prime\prime}1^\prime)
  \delta(2^{\prime\prime}2^\prime)}{n(1)M_m(1)n(2)M_m(2)g(12)}
\end{eqnarray}
This term describes two distinct particles.  The second term involves static correlations functions for 
more than two distinct particles.

In the spirit of a binary collision approximation, it would be reasonable to approximate both $W_{12}$ and 
$K_2$ by their first terms.  If this is done and the results are used in (\ref{eq:definitionQ12}), we get the 
following approximate result.
\begin{eqnarray}
\label{eq:approximateQ12}
\lefteqn{Q_{12}(1;1^\prime2^\prime)} \\ \nonumber 
  &\approx& \left(1+{\cal I}(1^\prime2^\prime)\right)\nabla_{\mathbf R}
          v_{MF}(12^\prime)\cdot\nabla_{\mathbf P}\delta(11^\prime) 
\end{eqnarray}
which is of the same form as the exact result but with the potential replaced by the potential of mean force.

Mazenko's theory contains quantities analogous to $W_{12}$ and $K_2$.  In the derivation of the MGBE, the 
contributions that correspond to more than two distinct particles are not included in each of these two 
functions. This would be analogous to using eq.~(\ref{eq:approximateQ12}) rather than (\ref{eq:exactQ12}).  
This is the origin of the fact that the final expression for the MGBE differs from the BCA in the way shown 
in eqs.~(\ref{eq:allV}) and (\ref{eq:MGBE-allV}). It is only in the factor of $Q_{12}$ that restricting 
attention to two particle effects leads to different results depending on whether the $Q$ vertices or the 
separate $W$ and $K$ functions are regarded as the quantities that are to be expressed in terms of two particle 
contributions.

\paragraph{Two comments.} 
1.\  The use of the exact result for $Q_{12}$ in the BCA is the origin of the fact that the BCA does not have 
the symmetry properties discussed in Appendix~\ref{appendix:sym}. The matter of whether an approximation 
satisfies the symmetry properties is determined by the relationships among all the approximations made rather 
than the accuracy of individual approximations.
2.\ This difference between the left vertices of the two theories (with the bare potential being replaced by a
factor of the potential of mean force) appears similar to differences noted in approximate kinetic theories of
hard spheres\cite{mazenko:73II,dorfman75} and of plasmas.\cite{gould75,gould77,boercker81,boercker82} 
The plasma case may be related to the present situation
but is more complicated in that both the left and right vertices are different in various versions of the theory.
The similarity to the hard sphere case is superficial, since the latter involves having different numbers of
factors of $g(d)$, the pair correlation function of hard spheres at contact, in the memory function. A factor of
the pair correlation function cannot arise from replacing the bare potential by the potential of mean force. The
differences in the hard sphere kinetic theories appear more closely related to
approximating $K_2$ by the first term in eq.~(\ref{eq:K2firstterm}) and then approximating 
the latter by its `disconnected' approximation
\begin{eqnarray*}
\lefteqn{K_2(1^{\prime\prime}2^{\prime\prime};1^\prime2^\prime)_{\mathrm{first\,term\,disconnected}}} \\*
 &=&\frac{\left(1+{\cal I}(1^{\prime\prime}2^{\prime\prime})\right)\delta(1^{\prime\prime}1^\prime)
  \delta(2^{\prime\prime}2^\prime)}{n(1)M_m(1)n(2)M_m(2)}
\end{eqnarray*}
which is equivalent to multiplying the approximate $K_2$ by a factor of the pair correlation function $g$.

\section{\label{appendix:numCalc}Numerical calculation of the matrix elements of the memory function and 
correlation functions} 
The matrix elements of the memory function can be expressed as an ensemble average over a two particle 
probability distribution function as follows
\begin{displaymath}
  \hat M_{\vec\mu\vec\nu}(\mathbf k,t) = 
       \langle W_{\vec\mu\vec\nu}(\mathbf k,t;12)\rangle\int_0^{r_c} dr\,4\pi r^2g(r)
\end{displaymath}
where $r_c$ is the cutoff distance associated with the potential governing the dynamics, that is $V_{22}$,
and the quantity being averaged is 
\begin{eqnarray*}
  \lefteqn{ W_{\vec\mu\vec\nu}(\mathbf k,t;12)=
  \nabla V_{Q_{21}}(\mathbf R)\cdot \nabla h_{\vec\nu}(\mathbf P_1)}  \\ 
&&\times  \nabla V_{Q_{12}}(|\mathbf R_{ab}|) 
\cdot [e^{-i\mathbf k\cdot\mathbf R_a}\nabla h_{\vec\mu}(\mathbf P_a)
     -e^{-i\mathbf k\cdot\mathbf R_b}\nabla h_{\vec\mu}(\mathbf P_b)]  
\end{eqnarray*}
The meaning of phase points $a$ and $b$ are described in Sct.~\ref{subsct:basisFuncExpan}. 
The first term in brackets gives the self part of the memory function, 
while the second term gives the distinct part. Both terms together constitute the total memory function.
The two particle distribution function required for calculation of the average
$\langle W_{\vec\mu\vec\nu}(\mathbf k,t;12)\rangle$
\begin{eqnarray*}
  P^{[2]}(12) \propto \rho M_m(\mathbf P_1) M_m(\mathbf P_2)\delta(\mathbf R_1)
       g(\mathbf R_2)\Theta(r_c-|\mathbf R_2|)
\end{eqnarray*}
The average of $W$ over this distribution is calculated by the two particle trajectory method of Ranganathan 
and Andersen.\cite{andersen:dkt5} In this method, the coordinates and momenta called $1,2$ are sampled from 
$P^{[2]}$ and used as initial conditions for a two particle trajectory calculation from which the $W$  
quantities are calculated as functions of time. For most of the approximations, the trajectories are
calculated using Hamiltonian mechanics of the two particles interacting via the potential $V_{22}$.  
For the  $K$ approximations, the dynamics are supplemented by an algorithm that includes the effect of 
the $\tilde L_{BGK}$ in eq.~(\ref{eq:Kapprox}). The algorithm gives each particle, at randomly chosen times, a 
new momentum chosen from a Boltzmann distribution, without changing its position. The time at which these 
interventions occur for each particle are uncorrelated with each other and occur with an average frequency
$\nu$ for each particle, where $\nu$ is given by eq.~(\ref{eq:nu}).

The matrix elements of the memory function can be divided into four components: self real, distinct 
real, self imaginary, and distinct imaginary parts. Each part is computed for each matrix element 
in a calculation that takes advantage of a number of properties derived from the Hermite polynomial basis 
functions, the two particle problem, and the known $t=0$ values of the matrix elements themselves.

The basis function $h_{\vec0}(\mathbf P)$ is a constant. Since the gradients of the Hermite polynomial 
functions with respect to momenta, not the Hermite functions themselves, 
appear in the expression for the matrix elements of the memory function, 
$\hat M_{\vec0\vec\nu}$ for all $\vec\nu$ and $\hat M_{\vec\mu\vec0}$ for all $\vec\mu$ are identically zero 
for all time and need not be computed numerically.

The two particle trajectory method can be mapped onto a highly symmetric two particle scattering problem. 
The symmetry relations of most concern are the following. 1.\ If $\mu_z+\nu_z$ is even, then the self 
and distinct imaginary parts are identically zero for all times.
2.\ If $\mu_z+\nu_z$ is odd, then the self and distinct real parts are identically zero for all times.
See Ref.~\onlinecite{jen:thesis} for the proofs of these properties. 

\begin{figure}
\epsfig{figure=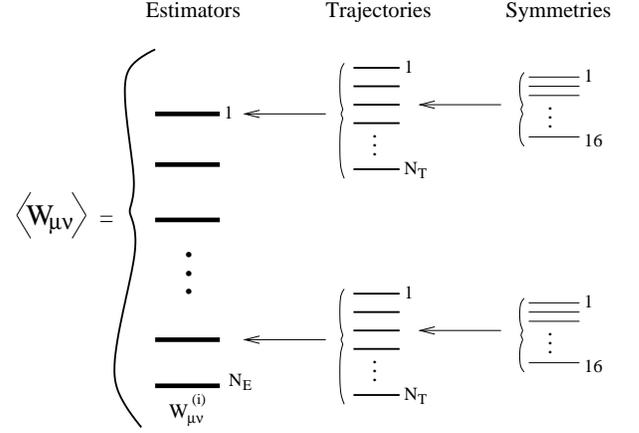,width=8cm}
\caption{A schematic representation of the averaging procedure for the matrix elements of the memory function. }
\label{fig:MEaverages}
\end{figure}

Figure~\ref{fig:MEaverages} is a schematic representation of the averaging procedure used to calculate the matrix elements of the memory
function. For each pair of indices $\vec\mu\vec\nu$, the following calculation scheme was used. 
1.\ A set of initial
conditions is randomly sampled from the two particle probability distribution $P^{[2]}$ and a molecular 
dynamics trajectory is computed by solving Hamilton's equations of motion. 
2.\ This single trajectory is used to generate fifteen symmetry
related two particle trajectories based on the $D_{4h}$ symmetry of the two particle system. The values of 
the various parts of $W_{\vec\mu\vec\nu}$ are calculated for each of the sixteen trajectories. This 
corresponds to generating numbers that appear in the rightmost column in
Fig.~\ref{fig:MEaverages}. The sixteen sets of numbers are averaged together to give the first type of 
average for the various parts of $W_{\vec\mu\vec\nu}$. The values are entered into the middle column of 
the figure. 
3.\ Steps 1 and 2 are repeated for a total of $N_T$ times. The $N_T$ results are averaged to give a second 
type of average, which is entered in the left column of the figure.
4.\ Steps 1-3 are repeated $N_E$ times to yield a set of $N_E$ values of the second type of average. 
These $N_E$ values are averaged together to give a third type of average, however they are not given equal 
weight as in steps 2 and 3. In this case, we use a constrained reweighting method, based on a maximum 
likelihood principle, that adjusts the weights assigned to each of the $N_E$ values such that the correct 
results are obtained for the calculated averages of the matrix elements at $t=0$. The result of this process 
is a set of values for the real and imaginary parts of the self and distinct parts of 
$\langle W_{\vec\mu\vec\nu}\rangle$. The values automatically satisfy the symmetry properties mentioned 
above, because of the way the symmetry related trajectories were used, and they automatically have the 
correct values at $t=0$, because of the reweighting method. The latter automatically reduces the 
statistical error for small nonzero times as well. Details of this method can be found in
Ref.~\onlinecite{jen:thesis}. The self parts of the average are used in eq.~(\ref{eq:Mmunu}) to give the 
desired matrix elements $\hat M^{(s)}_{\vec\mu\vec\nu}$.
5.\ The correlation functions of interest in this work, $\hat C^{(s)}_{\vec0\vec0}(\mathbf
k,t)$ and $\hat C^{(s)}_{\hat k\hat k}(\mathbf k,t)$, are computed by numerically solving the kinetic equation 
using the Euler method with the appropriate matrix elements of the memory function.

Steps 1-5 are repeated $N_s$ times to generate $N_s$ statistically independent results for the functions 
$\hat M_{\vec\mu\vec\nu}(t)$ and for $\hat C^{(s)}_{\vec0\vec0}(t)$ and $\hat C^{(s)}_{\hat k\hat k}(t)$. 
These $N_s$ results for each function are averaged, and statistical error bars are computed in the usual way. 
The results presented in this paper are these average quantities. The values of $N_T$, $N_E$ and $N_s$ used 
in this work are $N_T=100$, $N_E=1000$, and $N_s=10$.

\bibliography{refsDKT6}

\begin{thebibliography}{65}
\bibitem[{not({\natexlab{b}})}]{note2}
\bibinfo{note}{Most of the detailed work on the theory of liquids deals with
  the special case of atomic liquids. We shall focus on this case, but most of
  the ideas and results can be generalized to more complicated fluids. Here we
  are concerned only with the case of liquids with short ranged interatomic
  potentials and forces.}
\bibitem[{not({\natexlab{g}})}]{note9}
\bibinfo{note}{The BCA is in fact exact at $t=0$. See
  eqs.~(\ref{eq:MBCA-generalForm})--(\ref{eq:allV}).}
\end{thebibliography}

\end{document}